\documentclass{webofc}
\usepackage[varg]{txfonts}
\usepackage{latexsym,graphicx,epsfig,psfrag,here}
\usepackage{epstopdf}
\usepackage{amssymb,amsmath,amsxtra,amsfonts}
\usepackage{bm}

\usepackage{longtable}
\usepackage{multirow}
\usepackage{booktabs}
\usepackage{array}
\usepackage{wrapfig}
\usepackage{color}
\usepackage{titlesec}

\begin{document}

\title{$B_{s}\rightarrow K^{\ast 0}$ decay form factors from covariant confined quark model}

\author{\firstname{Aidos} \lastname{Issadykov}\inst{1,2,3}\fnsep\thanks{\email{issadykov@jinr.ru}} 
}

\institute{Joint Institute for Nuclear Research,141980 Dubna, Russia
  \and
 The Institute of Nuclear Physics,Ministry of Energy of the Republic of Kazakhstan, 050032 Almaty, Kazakhstan
\and
Al-Farabi Kazakh National University, SRI for Mathematics and Mechanics,050038 Almaty, Kazakhstan
}

\abstract{
 We evaluate $B_{s}\rightarrow K^{\ast 0}$ transition form factors in the full kinematical
region within the covariant  confined quark model. The calculated form factors can be used to calculate the $B_{s}\rightarrow K^{\ast 0}\mu^+ \mu^-$ rare decay branching ratio, which was recently measured by LHCb collaboration.
 }  
\maketitle

Last measurements of rare B-decays show deviations with respect to Standart Model
predictions\cite{Aaij:2013qta,Aaij:2013iag,Aaltonen:2011qs,Aaltonen:2011cn,Aaij:2013aln}. The $b\rightarrow s \ell^+ \ell^-$ and $b\rightarrow d \ell^+ \ell^-$ processes are forbidden at tree-level in Standart Model and sensitive to New Physics contributions in loops.The $b\rightarrow d$ transition is supressed than $b\rightarrow s$ due to CKM matrix elements. However it is interesting to study  decays proceeds via flavour-changing neutral-current (FCNC) transition.
The $b\rightarrow d$ transition decays observed by LHCb collaboration for $B^+\rightarrow \pi^{+}\mu^+ \mu^-$\cite{LHCb:2012de,Aaij:2015nea} and
$\Lambda^0\rightarrow \rho \pi^{-}\mu^+ \mu^-$~\cite{Aaij:2017ewm} decays.
Recently the LHCb Collaboration~\cite{Aaij:2018jhg} reported about the measurement of branching ratio of $B_{s}\rightarrow K^{\ast 0}\mu^+ \mu^-$ decay.

The $B_{s}\rightarrow K^{\ast 0}$ transition form factors  were studied in light-cone sum rule~\cite{Ball:2004rg,Straub:2015ica} and lattice QCD~\cite{Horgan:2015vla} techniques.
In view of these development, we calculate $B_{s}\rightarrow K^{\ast 0}$ form factors  within the covariant confined quark model(CCQM).

The covariant confined quark model\cite{Efimov:1988yd}
is an effective quantum field approach to hadronic interactions based on 
an interaction Lagrangian of hadrons interacting with their constituent quarks.
The value of the coupling constant follows form the compositeness 
condition~$Z_H=0$, where $Z_H$ is the wave function renormalization constant of the hadron. Matrix elements of the physical processes are generated by a set 
of quark loop diagrams according to the $1/N_c$ expansion. The ultraviolet 
divergences of the quark loops are regularized by including vertex functions 
for the hadron-quark vertices. These function also describe finite size 
effects related to the non-pointlike hadrons. The quark 
confinement \cite{Branz:2009cd} is built-in through an infrared cutoff on 
the upper limit of the scale integration 
to avoid the appearance of singularities in matrix elements. 
The infrared cutoff parameter $\lambda$ is universal for all processes. 
The  covariant confined quark model has limited number of parameters: the light and heavy constituent quark masses, 
the size  parameters which describe the size of the distribution 
of the constituent quarks inside the hadron and 
the infrared cutoff parameter $\lambda$. They are determined by a
fit to available experimental data.

In calculations we used next values of the model parameters which are shown in Eq.~(\ref{eq:fit}).

\begin{equation}
\def\arraystretch{1.5}
\begin{array}{ccccc|ccccc}
     m_{u/d}        &      m_s        &      m_c       &     m_b & \lambda  &   
 \Lambda_{B_s} & \Lambda_{K^{*0}} &   
 m_{B_s} & m_{K^{*0}}
\\\hline
 \ \ 0.241\ \   &  \ \ 0.428\ \   &  \ \ 1.67\ \   &  \ \ 4.68\ \   & 
\ \ 0.181\ \   &\ \ 2.05\ \   &\ \ 0.81\ \   &\ \ 5.367\ \   &\ \ 0.896\ \ & \ {\rm GeV} \\
\end{array}
\label{eq:fit}
\end{equation}

Below, we list the definitions of the dimensionless invariant transition
form factors together with the covariant quark model expressions that allow 
one to calculate them. We closely follow the notation used in our 
papers \cite{Dubnicka:2016nyy, Dubnicka:2015iwg}.

\begin{eqnarray}
&&
\langle 
V(p_2,\epsilon_2)_{[\bar q_3 q_2]}\,
|\,\bar q_2\, O^{\,\mu}\,q_1\, |\,P_{[\bar q_3 q_1]}(p_1)
\rangle 
\,=\,
\nonumber\\
&=&
N_c\, g_P\,g_V \!\! \int\!\! \frac{d^4k}{ (2\pi)^4 i}\, 
\widetilde\Phi_P\Big(-(k+w_{13})^2\Big)\,
\widetilde\Phi_V\Big(-(k+w_{23})^2\Big)
\nonumber\\
&\times&
{\rm tr} \biggl[ 
O^{\,\mu} \,S_1(k+p_1)\,\gamma^5\, S_3(k) \not\!\epsilon_2^{\,\,\dagger} \,
S_2(k+p_2)\, \biggr]
\nonumber\\
 & = &
\frac{\epsilon^{\,\dagger}_{\,\nu}}{m_1+m_2}\,
\Big( - g^{\mu\nu}\,P\cdot q\,A_0(q^2) + P^{\,\mu}\,P^{\,\nu}\,A_+(q^2)
       + q^{\,\mu}\,P^{\,\nu}\,A_-(q^2) 
\nonumber\\ 
&& + i\,\varepsilon^{\mu\nu\alpha\beta}\,P_\alpha\,q_\beta\,V(q^2)\Big),
\label{eq:PV}
\end{eqnarray}

\begin{eqnarray}
&&
\langle 
V(p_2,\epsilon_2)_{[\bar q_3 q_2]}\,
|\,\bar q_2\, (\sigma^{\,\mu\nu}q_\nu(1+\gamma^5))\,q_1\, |\,P_{[\bar q_3 q_1]}(p_1)
\rangle 
\,=\,
\nonumber\\
&=&
N_c\, g_P\,g_V \!\! \int\!\! \frac{d^4k}{ (2\pi)^4 i}\, 
\widetilde\Phi_P\Big(-(k+w_{13})^2\Big)\,
\widetilde\Phi_V\Big(-(k+w_{23})^2\Big)
\nonumber\\
&\times&
{\rm tr} \biggl[ 
(\sigma^{\,\mu\nu}q_\nu(1+\gamma^5))
\,S_1(k+p_1)\,\gamma^5\, S_3(k) \not\!\epsilon_2^{\,\,\dagger} \,S_2(k+p_2)\, 
\biggr]
\nonumber\\
 & = &
\epsilon^{\,\dagger}_{\,\nu}\,
\Big( - (g^{\mu\nu}-q^{\,\mu}q^{\,\nu}/q^2)\,P\cdot q\,a_0(q^2) 
       + (P^{\,\mu}\,P^{\,\nu}-q^{\,\mu}\,P^{\,\nu}\,P\cdot q/q^2)\,a_+(q^2)
\nonumber\\
&&
+ i\,\varepsilon^{\mu\nu\alpha\beta}\,P_\alpha\,q_\beta\,g(q^2)\Big).
\label{eq:PVT}
\end{eqnarray}

We use $P=p_1+p_2$ and $q=p_1-p_2$ and the on--shell conditions 
$\epsilon_2^\dagger\cdot p_2=0$,
$p_i^2=m_i^2$. Since there are three quark species involved in the transition,
we have introduced a two--subscript notation
$w_{ij}=m_{q_j}/(m_{q_i}+m_{q_j})$ $(i,j=1,2,3)$ such that $w_{ij}+w_{ji}=1$. 
The form factors defined in Eq.\,(\ref{eq:PVT}) satisfy the physical 
requirement $a_0(0)=a_+(0)$, which ensures that no kinematic singularity 
appears in the matrix element at $q^2=0$.  

The form factors are calculated in the full kinematical region of 
momentum transfer squared and results of our numerical calculations are with high accuracy approximated 
by the parametrization
\begin{equation}
F(q^2)=\frac{F(0)}{1-a s+b s^2}\,, \qquad s=\frac{q^2}{m_1^2}\,,
\label{eq:ff_approx}
\end{equation}
the relative error is less than 1$\%$.
The values of $F(0)$, $a$, and $b$ are listed  in Table~\ref{tab:apprff}.
\begin{table}[ht]
\caption{Parameters for the approximated form factors
in Eq.~(\ref{eq:ff_approx}).} 
\begin{center}
\begin{tabular}{c|rrrc|rrr}
\hline
&\qquad $A_0$ \qquad &\qquad $A_+$ \qquad  &\qquad $A_-$ \qquad & \qquad $V$ 
\qquad \quad 
&\qquad $a_0$ \qquad &\qquad $a_+$ \qquad &\qquad $g$ \qquad \\
\hline
$F(0)$ &  0.30   & 0.21 & $-0.23$ & 0.24 & 0.21    & 0.21 & 0.21 \\
$a$    &  -0.64   & -1.47 & -1.55    & -1.60 & -0.69    & -1.48 & -1.61 \\
$b$    & $-0.28$ & 0.44 & 0.52   & 0.56 & $-0.23$ & 0.45 & 0.57 \\
\hline
\end{tabular}
\label{tab:apprff} 
\end{center}
\end{table}

\begin{figure*}[htbp]
\begin{center}
\epsfig{figure=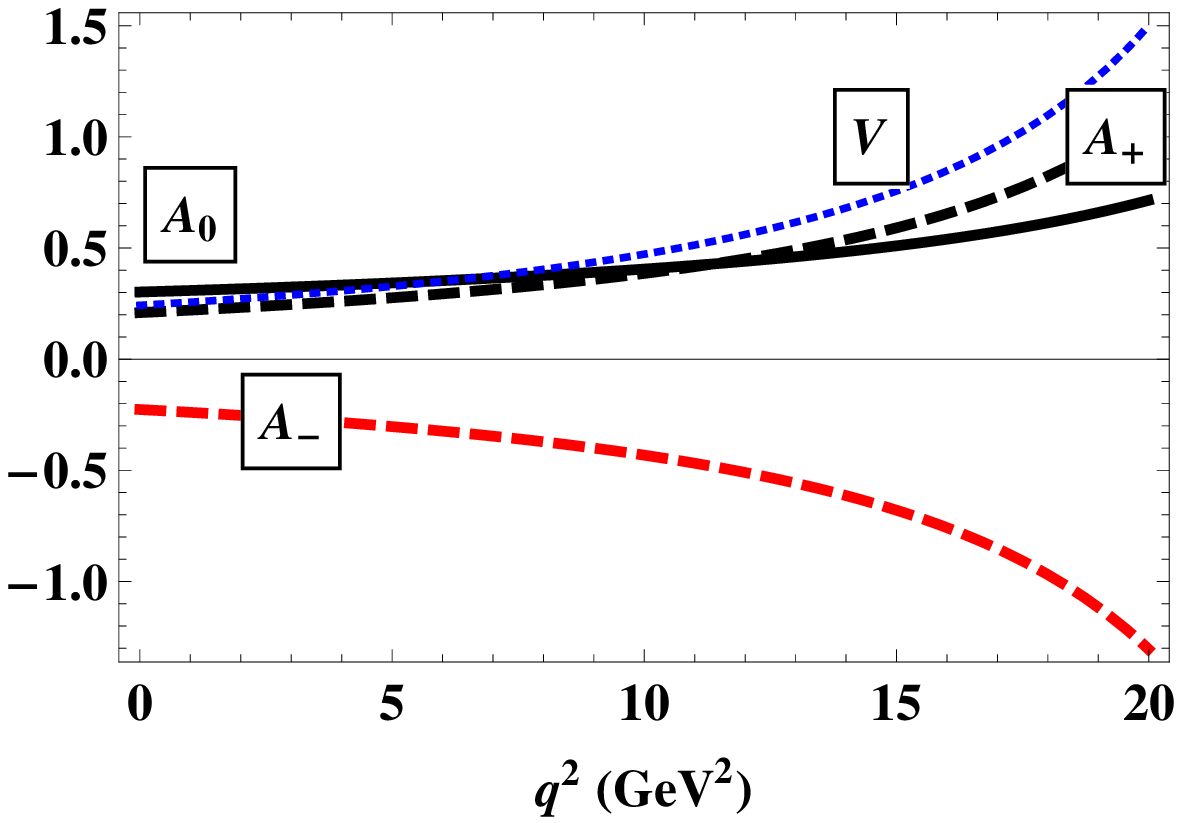,scale=0.6}
\hspace*{.25cm}
\epsfig{figure=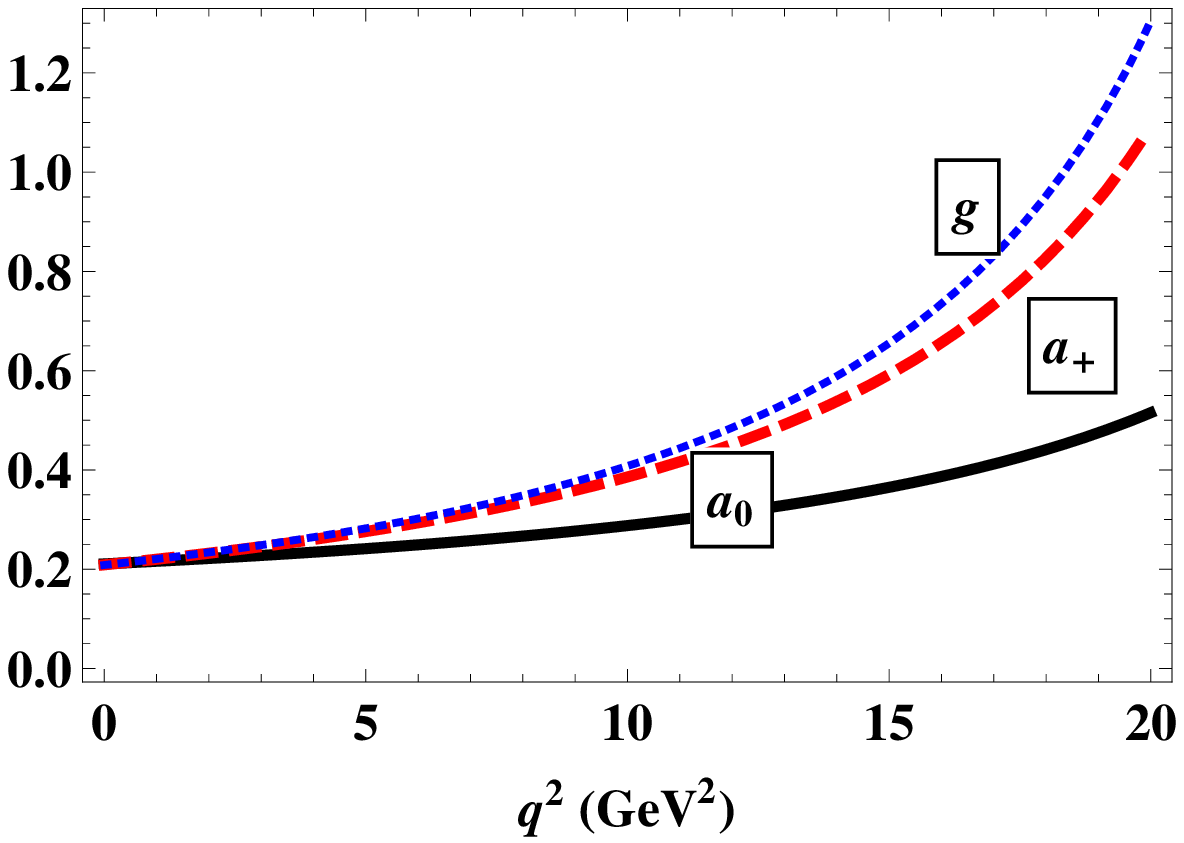,scale=0.6}
\caption{The  $q^{2}$-dependence of the vector and axial form factors
(upper plot) and tensor form factors (lower plot) for $B_{s}\rightarrow K^{\ast 0}$ decay.
\label{fig:BsKv-FF}
}
\end{center}
\end{figure*}

The curves are depicted  in Fig.~\ref{fig:BsKv-FF}.

 The obtained errors of the fitted parameters were of the order of $10\%$. 
 Indeed it follows that form factors at $q^{2}=0$ were calculated with $10\%$
 uncertainties. This implies at least $10\%$ uncertainty in form factors
 in the full kinematical region of momentum transfer squared.

For reference it is useful to relate the above form factors  
to those used, e.g., in Ref.\,\cite{Ball:2004rg} 
(we denote them by the superscript $^c$).
The relations read 
\begin{eqnarray}
A_0 &=& \frac{m_1 + m_2}{m_1 - m_2}\,A_1^c\,, \qquad 
A_+ = A_2^c\,,
\nonumber\\
A_- &=&  \frac{2m_2(m_1+m_2)}{q^2}\,(A_3^c - A_0^c)\,, \qquad
V = V^c\,, 
\nonumber\\[1.2ex]
a_0 &=& T_2^c\,, \qquad g = T_1^c\,, \qquad
a_+  =  T_2^c + \frac{q^2}{m_1^2-m_2^2}\,T_3^c\,.
\label{eq:new-ff}
\end{eqnarray}
We note in addition that the form factors (\ref{eq:new-ff}) satisfy
the constraints
\begin{eqnarray}
 A_0^c(0) &=& A_3^c(0) 
\nonumber\\
2m_2A_3^c(q^2) &=& (m_1+m_2) A_1^c(q^2) -(m_1-m_2) A_2^c(q^2)\,.
\end{eqnarray}

Since $a_0(0)=a_+(0)=g(0)$ we display in Table~\ref{tab:ff-comparison}
the form factors
$A_0^c(0)=(m_1-m_2)[A_0(0)-A_{+}(0)]/(2m_2)$,
$A_1^c(0)=A_0(0)(m_1-m_2)/(m_1+m_2)$,
$A_2^c(0)=A_+(0)$,
$T_1^c(0)=g(0)$ and 
$T_3^c(0)=\lim_{\,q^2 \to 0}(m_1^2-m_2^2)(a_{+}-a_0)/q^2$
obtained in our model and compare them with
those from light-cone sum rule~\cite{Ball:2004rg}.

\begin{table}
\caption{The form factors at  maximum recoil $q^2=0$.}
\label{tab:ff-comparison}
\begin{tabular}{cllllll}
\toprule
     & $V^c(0)$ & $A_0^c(0)$ &$A_1^c(0)$ &$A_2^c(0)$ &$T_1^c(0)$ &$T_3^c(0)$ \\
\hline

CCQM & $0.24\pm 0.02$ & $0.18\pm 0.02$
 & $0.21\pm 0.02$ & $0.21\pm 0.02$ & $0.21\pm 0.02$ & $0.14\pm 0.01 $\\

Ref.~\cite{Ball:2004rg}& 0.31 &0.36  & 0.23 & 0.18 & 0.26 &0.14 \\
\end{tabular}
\end{table}

\section*{Acknowledgment}
We thank Prof. Mikhail A. Ivanov for the continuous support through out
this work and  for useful discussions of some aspects.

The work has been carried out under financial support of the
Program of the Ministry of Education and Science of the
Republic of Kazakhstan IRN  number AP05132978.

Author A. Issadykov is grateful for the support by the JINR, grant number 
18-302-03.


\begin{thebibliography}{99}

\bibitem{Aaij:2013qta}
  R.~Aaij {\it et al.} [LHCb Collaboration],
  Phys.\ Rev.\ Lett.\  {\bf 111} (2013) 191801
  doi:10.1103/PhysRevLett.111.191801
  [arXiv:1308.1707 [hep-ex]].

\bibitem{Aaij:2013iag}
  R.~Aaij {\it et al.} [LHCb Collaboration],
  JHEP {\bf 1308} (2013) 131
  doi:10.1007/JHEP08(2013)131
  [arXiv:1304.6325 [hep-ex]].

\bibitem{Aaltonen:2011qs}
  T.~Aaltonen {\it et al.} [CDF Collaboration],
  Phys.\ Rev.\ Lett.\  {\bf 107} (2011) 201802
  doi:10.1103/PhysRevLett.107.201802
  [arXiv:1107.3753 [hep-ex]].

\bibitem{Aaltonen:2011cn}
  T.~Aaltonen {\it et al.} [CDF Collaboration],
  Phys.\ Rev.\ Lett.\  {\bf 106} (2011) 161801
  doi:10.1103/PhysRevLett.106.161801
  [arXiv:1101.1028 [hep-ex]].

\bibitem{Aaij:2013aln}
  R.~Aaij {\it et al.} [LHCb Collaboration],
  JHEP {\bf 1307} (2013) 084
  doi:10.1007/JHEP07(2013)084
  [arXiv:1305.2168 [hep-ex]].
  
\bibitem{LHCb:2012de}
  R.~Aaij {\it et al.} [LHCb Collaboration],
  JHEP {\bf 1212} (2012) 125
  doi:10.1007/JHEP12(2012)125
  [arXiv:1210.2645 [hep-ex]].
  
\bibitem{Aaij:2015nea}
  R.~Aaij {\it et al.} [LHCb Collaboration],
  JHEP {\bf 1510} (2015) 034
  doi:10.1007/JHEP10(2015)034
  [arXiv:1509.00414 [hep-ex]].

\bibitem{Aaij:2017ewm}
  R.~Aaij {\it et al.} [LHCb Collaboration],
  JHEP {\bf 1704} (2017) 029
  doi:10.1007/JHEP04(2017)029
  [arXiv:1701.08705 [hep-ex]].

\bibitem{Aaij:2018jhg}
  R.~Aaij {\it et al.} [LHCb Collaboration],
  JHEP {\bf 1807} (2018) 020
  doi:10.1007/JHEP07(2018)020
  [arXiv:1804.07167 [hep-ex]].
    
\bibitem{Ball:2004rg}
  P.~Ball and R.~Zwicky,
  Phys.\ Rev.\ D {\bf 71} (2005) 014029
  doi:10.1103/PhysRevD.71.014029
  [hep-ph/0412079].

\bibitem{Straub:2015ica}
  A.~Bharucha, D.~M.~Straub and R.~Zwicky,
  JHEP {\bf 1608} (2016) 098
  doi:10.1007/JHEP08(2016)098
  [arXiv:1503.05534 [hep-ph]].

\bibitem{Horgan:2015vla}
  R.~R.~Horgan, Z.~Liu, S.~Meinel and M.~Wingate,
  PoS LATTICE {\bf 2014} (2015) 372
  [arXiv:1501.00367 [hep-lat]].

\bibitem{Efimov:1988yd}
  G.~V.~Efimov and M.~A.~Ivanov,
  Int.\ J.\ Mod.\ Phys.\ A {\bf 4} (1989) 2031.
  doi:10.1142/S0217751X89000832

\bibitem{Branz:2009cd}
  T.~Branz, A.~Faessler, T.~Gutsche, M.~A.~Ivanov, J.~G.~K\"orner and V.~E.~Lyubovitskij,
  Phys.\ Rev.\ D {\bf 81} (2010) 034010
  doi:10.1103/PhysRevD.81.034010
  [arXiv:0912.3710 [hep-ph]]

\bibitem{Dubnicka:2016nyy}
  S.~Dubni\v{c}ka, A.~Z.~Dubni\v{c}ová, A.~Issadykov, M.~A.~Ivanov, A.~Liptaj and S.~K.~Sakhiyev,
  Phys.\ Rev.\ D {\bf 93} (2016) no.9,  094022
  doi:10.1103/PhysRevD.93.094022
  [arXiv:1602.07864 [hep-ph]].

\bibitem{Dubnicka:2015iwg}
  S.~Dubni\v{c}ka, A.~Z.~Dubni\v{c}ková, N.~Habyl, M.~A.~Ivanov, A.~Liptaj and G.~S.~Nurbakova,
  Few Body Syst.\  {\bf 57} (2016) no.2,  121
  doi:10.1007/s00601-015-1034-4
  [arXiv:1511.04887 [hep-ph]].

  
\end{thebibliography}
\end{document}